\documentclass[a4paper]{amsart}
\RequirePackage{amsmath}
\RequirePackage{bm}
\RequirePackage{amssymb}
\RequirePackage{upref}
\RequirePackage{amsthm}
\RequirePackage{enumerate}
\RequirePackage{pb-diagram}
\RequirePackage{amsfonts}
\RequirePackage[mathscr]{eucal}
\RequirePackage{verbatim}
\RequirePackage{xr}
\RequirePackage{graphicx}
\usepackage{calc}
\usepackage{xspace}
\RequirePackage{color}
\RequirePackage{ifthen}
\RequirePackage{fancybox}

\newcommand{\cf}{cf.\@\xspace}
\newcommand{\resp}{resp.\@\xspace}

\newcommand{\al}{\alpha}
\newcommand{\bet}{\beta}
\newcommand{\ga}{\gamma}
\newcommand{\de}{\delta }
\newcommand{\e}{\epsilon}

\newcommand{\f}{\varphi}

\newcommand{\lam}{\lambda}

\newcommand{\n}{\nu}
\newcommand{\om}{\omega}

\newcommand{\s}{\sigma}
\newcommand{\x}{\xi}

\newcommand{\C}{\varGamma}
\newcommand{\D}{\varDelta}
\newcommand{\F}{\varPhi}
\newcommand{\Lam}{\varLambda}
\newcommand{\Om}{\varOmega}

\newcommand{\di}[1]{#1\nobreakdash-\hspace{0pt}dimensional}%\di n

\newcommand{\fv}[2]{#1\hspace{0pt}_{|_{#2}}}

\newcommand{\so}{{\mc S_0}}

\newcommand{\const}{\tup{const}}

\newcommand{\msp[1]}[1]{\mspace{#1mu}}

\newcommand{\R}[1][n+1]{{\protect\mathbb R}^{#1}}

\newcommand{\Cc}{{\protect\mathbb C}}

\newcommand{\N}{{\protect\mathbb N}}

\newcommand{\eR}{\stackrel{\lower1ex \hbox{\rule{6.5pt}{0.5pt}}}{\msp[3]\R[]}}
\newcommand{\eN}{\stackrel{\lower1ex \hbox{\rule{6.5pt}{0.5pt}}}{\msp[1]\N}}
\newcommand{\eO}{\stackrel{\lower1ex \hbox{\rule{6pt}{0.5pt}}}{\msc O}}
\newcommand{\mf}[1]{\mathfrak {#1}}

\DeclareMathOperator{\graph}{graph}

\DeclareMathOperator{\Ad}{Ad}

\newcommand\ra{\rightarrow}

\newcommand\pde[2]{\frac {\partial#1}{\partial#2}}
\newcommand\dde[2]{\frac {\de#1}{\de#2}}

\newcommand{\un}{\infty}
\newcommand{\A}{\forall}

\newcommand{\set}[2]{\{\,#1\colon #2\,\}}
\newcommand{\uu}{\cup}
\newcommand{\ii}{\cap}
\newcommand{\uuu}{\bigcup}

\newcommand{\uud}{ \stackrel{\lower 1ex \hbox {.}}{\uu}}
\newcommand{\uuud}[1]{ \stackrel{\lower 1ex \hbox {.}}{\uuu_{#1}}}
\newcommand\su{\subset}
\newcommand\Su{\Subset}

\newcommand{\sminus}[1][28]{\raise 0.#1ex\hbox{$\scriptstyle\setminus$}}

\newcommand{\wed}{\wedge}

\newcommand{\abs}[1]{\lvert#1\rvert}

\newcommand{\norm}[1]{\lVert#1\rVert}

\newcommand{\spd}[2]{\protect\langle #1,#2\protect\rangle}

\newcommand\ch[3]{\varGamma_{#1#2}^#3}
\newcommand\cha[3]{{\bar\varGamma}_{#1#2}^#3}

\newcommand{\riem}[4]{R_{#1#2#3#4}}
\newcommand{\riema}[4]{{\bar R}_{#1#2#3#4}}

\newcommand{\tit}{\textit}

\newcommand{\tup}{\textup}% text upright

\newcommand{\mc}{\protect\mathcal}
\newcommand{\msc}{\protect\mathscr}

\providecommand{\bysame}{\makebox[3em]{\hrulefill}\thinspace}

\newcommand{\bt}{\begin{thm}}
\newcommand{\bl}{\begin{lem}}
\newcommand{\bc}{\begin{cor}}
\newcommand{\bd}{\begin{definition}}
\newcommand{\bpp}{\begin{prop}}
\newcommand{\br}{\begin{rem}}
\newcommand{\bn}{\begin{note}}
\newcommand{\be}{\begin{ex}}
\newcommand{\bes}{\begin{exs}}
\newcommand{\bb}{\begin{example}}
\newcommand{\bbs}{\begin{examples}}
\newcommand{\ba}{\begin{axiom}}
\newcommand{\bas}{\begin{assumption}}

\newcommand{\et}{\end{thm}}
\newcommand{\el}{\end{lem}}
\newcommand{\ec}{\end{cor}}
\newcommand{\ed}{\end{definition}}
\newcommand{\epp}{\end{prop}}
\newcommand{\er}{\end{rem}}
\newcommand{\en}{\end{note}}
\newcommand{\ee}{\end{ex}}
\newcommand{\ees}{\end{exs}}
\newcommand{\eb}{\end{example}}
\newcommand{\ebs}{\end{examples}}
\newcommand{\ea}{\end{axiom}}
\newcommand{\eas}{\end{assumption}}

\newcommand{\bp}{\begin{proof}}
\newcommand{\ep}{\end{proof}}
\newcommand{\eps}{\renewcommand{\qed}{}\end{proof}}

\newcommand{\bal}{\begin{align}}

\newcommand{\bi}[1][1.]{\begin{enumerate}[\upshape #1]}
\newcommand{\bia}[1][(1)]{\begin{enumerate}[\upshape #1]}
\newcommand{\bin}[1][1]{\begin{enumerate}[\upshape\bfseries #1]}
\newcommand{\bir}[1][(i)]{\begin{enumerate}[\upshape #1]}
\newcommand{\bic}[1][(i)]{\begin{enumerate}[\upshape\hspace{2\cma}#1]}
\newcommand{\bis}[2][1.]{\begin{enumerate}[\upshape\hspace{#2\parindent}#1]}
\newcommand{\ei}{\end{enumerate}}

\newcommand\ndots{\raise 0.47ex \hbox {,}\hskip0.06em\cdots %
     \raise 0.47ex \hbox {,}\hskip0.06em} 

\newcommand{\q}{\quad}
\newcommand{\qq}{\qquad}

\newcommand{\hp}{\hphantom}

\newcommand\nd{\noindent}

\newskip\Csmallskipamount                                                
\Csmallskipamount=\smallskipamount
\newskip\Cmedskipamount
\Cmedskipamount=\medskipamount
\newskip\Cbigskipamount
\Cbigskipamount=\bigskipamount

\newcommand\cvs{\vspace\Csmallskipamount}   
\newcommand\cvm{\vspace\Cmedskipamount}

\newskip\csa
\csa=\smallskipamount

\newskip\cma
\cma=\medskipamount

\newskip\cba
\cba=\bigskipamount

\newdimen\spt
\newcommand\citem{\cvs\advance\itemno by
1{(\romannumeral\the\itemno})\hskip3pt}
\newcommand{\bitem}{\cvm\nd\advance\itemno by
1{\bf\the\itemno}\hspace{\cma}}

\newcount\itemno
\newcommand{\lae}[1]{\label{E:#1}}
\newcommand{\lat}[1]{\label{T:#1}}
\newcommand{\lal}[1]{\label{L:#1}}
\newcommand{\lad}[1]{\label{D:#1}}

\newcommand{\rt}[1]{Theorem~\ref{T:#1}}
\newcommand{\rl}[1]{Lemma~\ref{L:#1}}

\newcommand{\re}[1]{\eqref{E:#1}}

\newcommand{\fre}[1]{\eqref{E:#1} on page~\tup{\pageref{E:#1}}}

\newskip\thmskip
\thmskip=\parindent

\newskip\hsk
\newenvironment{hinw}{\labelsep=0pt\begin{list}{}{\labelsep=0pt\itemindent=0pt\labelwidth=0pt\leftmargin=\parindent\rightmargin=0pt\partopsep=\cba}%
\item\it\nopagebreak\nopagebreak}%
{\end{list}}

\newcommand\bh{\begin{hinw}}
\newcommand{\eh}{\end{hinw}}

\newtheoremstyle{normal}% name
  {\cba}%      Space above, empty = `usual value'
  {\cba}%      Space below
  {}% Body font
  {\thmskip}%Indent amount (empty = no indent, \parindent = para indent)
  {\bfseries}% Thm head font
  {.}%        Punctuation after thm head
  {\hsk}%     Space after thm head: " " = normal interword space;
  {}% Thm head spec

\newtheoremstyle{abschnitt}% name
  {\cba}%      Space above, empty = `usual value'
  {\cba}%      Space below
  {}% Body font
  {\thmskip}% Indent amount (empty = no indent, \parindent = para indent)
  {\bfseries}% Thm head font
  {.}%        Punctuation after thm head
  {\hsk}%     Space after thm head: " " = normal interword space;
  {}% Thm head spec

\newtheoremstyle{italic}% name
  {\cba}%      Space above, empty = `usual value'
  {\cba}%      Space below
  {\itshape}% Body font
  {\thmskip}%  Indent amount (empty = no indent, \parindent = para indent)
  {\bfseries}% Thm head font
  {.}%        Punctuation after thm head
  {\hsk}%     Space after thm head: " " = normal interword space;
  {}% Thm head spec

\newtheoremstyle{aufgaben}% name
  {\cba}%      Space above, empty = `usual value'
  {\cba}%      Space below
  {}% Body font
  {}%         Indent amount (empty = no indent, \parindent = para indent)
  {\normalsize\bfseries}% Thm head font
  {.}%        Punctuation after thm head
  {\hsk}%     Space after thm head: " " = normal interword space;
  {}% Thm head spec

\newtheoremstyle{break}% name
  {\cba}%      Space above, empty = `usual value'
  {\cba}%      Space below
  {\itshape}% Body font
  {}%         Indent amount (empty = no indent, \parindent = para indent)
  {\bfseries}% Thm head font
  {.}%        Punctuation after thm head
  {\newline}% Space after thm head: \newline = linebreak
  {}%         Thm head spec

\swapnumbers
\theoremstyle{italic}
\newtheorem{thm}[subsection]{Theorem}
\newtheorem{lem}[subsection]{Lemma}
\newtheorem{prop}[subsection]{Proposition}
\newtheorem{cor}[subsection]{Corollary}

\theoremstyle{normal}
\newtheorem{rem}[subsection]{Remark}
\newtheorem{definition}[subsection]{Definition}
\newtheorem{example}[subsection]{Example}
\newtheorem{examples}[subsection]{Examples}
\newtheorem{ex}[subsection]{Exercise}
\newtheorem{note}[subsection]{}
\newtheorem{axiom}[subsection]{Axiom}
\newtheorem{assumption}[subsection]{Assumption}

\theoremstyle{aufgaben}
\newtheorem{exs}[subsection]{Exercises}

\swapnumbers

\numberwithin{equation}{section}
\numberwithin{figure}{section}

\newenvironment{textequation}[1][0.8]
{\begin{equation}
\begin{aligned}
\begin{minipage}{#1\linewidth}}
{\end{minipage}
\end{aligned}
\end{equation}
\ignorespacesafterend}

\newcommand{\btext}{\begin{textequation}}
\newcommand{\etext}{\end{textequation}}

\def\hinweis{\@startsection{subsection}{2}%
 \z@{0.7\linespacing\@plus 0.5\linespacing}{0.7\linespacing}%
{\normalfont\itshape\indent}}

\newcounter{hours}\newcounter{minutes}
\newcommand{\printtime}{%
\setcounter{hours}{\time/60}%
\setcounter{minutes}{\time-\value{hours}*60}%
\ifthenelse{\value{minutes}<10}{\thehours :0\theminutes}{\thehours:\theminutes}}

\usepackage[german,english]{babel}
\usepackage{graphicx}
\RequirePackage{amsmath}
\RequirePackage{bm}
\RequirePackage{amssymb}
\RequirePackage{upref}
\RequirePackage{amsthm}
\RequirePackage{enumerate}
\RequirePackage{pb-diagram}
\RequirePackage{amsfonts}
\RequirePackage[mathscr]{eucal}
\RequirePackage{verbatim}
\RequirePackage{xr}
\RequirePackage{graphicx}
\usepackage{calc}
\usepackage{xspace}
\usepackage{dsfont}

\makeatletter
\RequirePackage{color}
\newcommand{\ann}[1]{\renewcommand{\@makefnmark}{\mbox{$^{\color{red}{\@thefnmark}}$}}%
\footnote {#1}}
\makeatother

\RequirePackage{upref}
\RequirePackage{amsthm}
\RequirePackage{enumerate}%\begin{enumerate}[(i)]
\usepackage[mathscr]{eucal}
\usepackage{xr-hyper}

\usepackage{calc}

\newlength{\oddsidemarginlength}
\newlength{\topmarginlength}

\newcounter{numberoflines}
\newcounter{tempcc}
\oddsidemargin=\oddsidemarginlength
\evensidemargin=\oddsidemargin
\topmargin=\topmarginlength
\usepackage[colorlinks=true,linkcolor=blue,citecolor=blue,urlcolor=blue]{hyperref}  

\begin{document}

\flushbottom

%\larger[1]
%\frontmatter

\title[Gravity interacting with a Yang-Mills field]{A unified field theory II: Gravity interacting with a Yang-Mills  and Higgs field}

% author one information
\author{Claus Gerhardt}
\address{Ruprecht-Karls-Universit\"at, Institut f\"ur Angewandte Mathematik,
Im Neuenheimer Feld 294, 69120 Heidelberg, Germany}
%\curraddr{}
\email{\href{mailto:gerhardt@math.uni-heidelberg.de}{gerhardt@math.uni-heidelberg.de}}
\urladdr{\href{http://www.math.uni-heidelberg.de/studinfo/gerhardt/}{http://www.math.uni-heidelberg.de/studinfo/gerhardt/}}
%\thanks{This work was supported by the DFG}

% author two information
%\author{}
%\address{}
%\curraddr{}
%\email{}
%\thanks{}
%
\subjclass[2000]{83,83C,83C45}
\keywords{unified field theory, quantization of gravity, quantum gravity, Yang-Mills fields, mass gap}
\date{\today}
%
% at present the "communicated by" line appears only in ERA and PROC
%\commby{}

%\dedicatory{}

\begin{abstract} 
We quantize the interaction of gravity with a Yang-Mills and Higgs field using canonical quantization. Similar to the approach in a previous paper we discard the Wheeler-DeWitt equation and express the Hamilton constraint by the evolution equation of the mean curvature of the hypersurfaces in the foliation defined by the Hamiltonian setting. Expressing the time derivative of the mean curvature with the help of the Poisson brackets the canonical quantization of this equation leads to a wave equation in $Q=(0,\infty)\times \so$, where $\so$ is one of the Cauchy hypersurfaces in the Hamiltonian setting. The wave equation describes the interaction of an arbitrary Riemannian metric in $\so$ and a given Yang-Mills and Higgs field. If the metric is complete $Q$ is globally hyperbolic. In case $\so$ is compact we also prove a spectral resolution of the wave equation and establish sufficient conditions guaranteeing a mass gap.
\end{abstract}

\maketitle

\tableofcontents

\setcounter{section}{0}
\section{Introduction}
The quantization of gravity interacting with Yang-Mills and Higgs fields poses no additional greater challenges---at least in principle. The number of variables will be increased, the combined Hamiltonian is the sum of several individual Hamiltonians, and, since gravity is involved, we have the Hamilton constraint as a side condition. Deriving the Einstein equations by a Hamiltonian setting requires a global time function $x^0$ and foliation of spacetime by its level hypersurfaces. Thus, we consider a spacetime $N=N^{n+1}$ with metric $(\bar g_{\al\bet})$, $0\le \al.\bet\le n$, assuming the existence of a global time function $x^0$ which will also define the time coordinate. Furthermore, we only consider metrics that can be split by the time function, i.e., the metrics can be expressed in the form
\begin{equation}\lae{1.1}
d\bar s^2=-w^2(dx^0)^2+g_{ij}dx^idx^j,
\end{equation}
where $w>0$ is a smooth function and $g_{ij}(x^0,x)$ are Riemannian metrics. Let
\begin{equation}\lae{1.2}
M(t)=\{x^0=t\}, \qq t\in x^0(N)\equiv I,
\end{equation}
be the coordinate slices, then the $g_{ij}$ are the induced metrics. Moreover, let $\mc G$ be a compact, semi-simple, connected Lie group with Lie algebra $\mf g$, and let 
\begin{equation}
E_1=(N,\mf g, \pi, \Ad(\mc G))
\end{equation}
be the corresponding adjoint bundle with bas space $N$. Then we consider the functional
\begin{equation}
J=\int_N(\bar R-2\Lam)+\int_N(\al_1 L_{YM}+\al_2L_H),
\end{equation}
where the $\al_i$, $i=1,2$, are positive coupling constants, $\bar R$ the scalar curvature, $\Lam$ a cosmological constant, $L_{YM}$ the energy of a connection in $E_1$ and $L_H$ the energy of a Higgs field with values in $\mf g$.  The integration over $N$ is to be understood symbolically, since we shall always integrate over an open precompact subset $\tilde\Om\su N$.

In a former paper \cite{cg:uqtheory} we already considered a canonical quantization of the above action and proved that it will be sufficient to only consider connections $A^a_\mu$ satisfying the Hamilton gauge
\begin{equation}
A^a_0=0,
\end{equation}
thereby eliminating the Gau{\ss} constraint, such that the only remaining constraint is the Hamilton constraint, \cf \cite[Theorem 2.3]{cg:uqtheory}.

Using the $ADM$ partition \re{1.2} of $N$, \cf \cite{adm:old}, such that
\begin{equation}
N=I\times \so, 
\end{equation}
where $\so$ is the Cauchy hypersurface $M(0)$ and applying canonical quantization we obtained a Hamilton operator $\mc H$ which was a normally hyperbolic operator in a fiber bundle $E$ with base space $\so$ and fibers
\begin{equation}
F(x)\times (\mf g\otimes T^{0,1}_x(\so))\times \mf g,\qq x\in \so,
\end{equation}
where $F(x)$ is the space of Riemannian metrics. We quantized the action by looking at the Wheeler-DeWitt equation in this bundle. The fibers of $E$ are equipped with a Lorentzian metric such that they are globally hyperbolic and the transformed Hamiltonian $\mc H$, which is now a hyperbolic operator, is a normally hyperbolic operator acting only in the fibers. 

The Wheeler-DeWitt equation has the form
\begin{equation}
\mc Hu=0,
\end{equation}
with $u\in C^\un(E,\Cc)$ and we defined with the help of the Green's operator a symplectic vector space and a corresponding Weyl system.

The Wheeler-DeWitt equation seems to be the obvious quantization of the Hamilton condition. However, $\mc H$ acts only in the fibers and not in the base space which is due to the fact that the derivatives are only ordinary covariant derivatives and not functional derivatives, though they are supposed to be functional derivatives, but this property is not really invoked when a functional derivative is applied to $u$, since the result is the same as  applying a partial derivative.

Therefore, we shall  discard the Wheeler-DeWitt equation and express the Hamilton condition differently by looking at the evolution equation of the mean curvature of the foliation hypersurfaces $M(t)$ and implementing the Hamilton condition on the right-hand side of this evolution equation. The left-hand side, a time derivative, we shall express with the help of the Poisson brackets. After canonical quantization the Poisson brackets  become a commutator and now we can employ the fact that the derivatives are functional derivatives, since we have to differentiate the scalar curvature of a metric. As a result we obtain an elliptic differential operator in the base space, the main part of which is the Laplacian of the metric.

On the right-hand side of the evolution equation  the interesting term is $H^2$, the square of the mean curvature. It will be transformed to a second time derivative and will be the only remaining derivative with respect to a fiber variable, since the differentiations with respect to the other variables cancel each other.

The resulting quantized equation is then a wave equation
\begin{equation}\lae{1.9}
\begin{aligned}
&\frac1{32}\frac {n^2}{n-1}\Ddot u-(n-1)t^{2-\frac4n}\D u-\frac n2t^{2-\frac4n}Ru+\al_1\frac n8 t^{2-\frac4n}F_{ij}F^{ij}u\\
&+\al_2 \frac n4 t^{2-\frac4n}\ga_{ab}\s^{ij}\F^a_i\F^b_iu+\al_2\frac n2 m t^{2-\frac4n}V(\F)u+nt^2\Lam u=0,
\end{aligned}
\end{equation}
in a globally hyperbolic spacetime
\begin{equation}
Q=(0,\un)\times \so
\end{equation}
describing the interaction of a given complete Riemannian metric $\s_{ij}$ in $\so$ with a  given Yang-Mills  and Higgs field; $V$ is the potential of the Higgs field and $m$ a positive constant. The existence of the time variable, and its range, is due to the Lorentzian metric in the fibers of $E$.
\br
For the results and arguments in this paper it is completely irrelevant that the values of the Higgs field $\F$ lie in a Lie algebra, i.e., $\F$ could also be just an arbitrary scalar field, or we could consider a Higgs field as well as an another arbitrary scalar field. Hence, let us stipulate that the Higgs field could also be just an arbitrary scalar field. It will later be used to produce a mass gap simply by interacting with the gravitation  ignoring the Yang-Mills field.
\er
If $\so$ is compact we also prove a spectral resolution of equation \re{1.9} by first considering a stationary version of the hyperbolic equation, namely, the elliptic eigenvalue equation
\begin{equation}\lae{1.11}
\begin{aligned}
&-(n-1)\D v-\frac n2Rv+\al_1\frac n8 F_{ij}F^{ij}v\\
&+\al_2 \frac n4 \ga_{ab}\s^{ij}\F^a_i\F^b_iv+\al_2\frac n2 m V(\F)v=\mu v.
\end{aligned}
\end{equation}
It has countably many solutions $(v_i,\mu_i)$ such that
\begin{equation}
\mu_0<\mu_1\le \mu_2\le \cdots,
\end{equation}
\begin{equation}
\lim \mu_i=\un.
\end{equation}
Let $v$ be an eigenfunction with eigenvalue $\mu>0$, then we  look at solutions of \re{1.9} of the form
\begin{equation}
u(x,t)=w(t) v(x).
\end{equation}
$u$ is then a solution of \re{1.9} provided $w$ satisfies the implicit eigenvalue equation
\begin{equation}\lae{1.15}
-\frac1{32}\frac{n^2}{n-1}\Ddot w-\mu t^{2-\frac4n}w-nt^2\Lam w=0,
\end{equation}
where $\Lam$ is the eigenvalue.

This eigenvalue problem we also considered in a previous paper and proved that it has countably many solutions $(w_i,\Lam_i)$ with finite energy, i.e.,
\begin{equation}
\int_0^\un\{\abs{\dot w_i}^2+(1+t^2+\mu t^{2-\frac4n})\abs{w_i}^2\}<\un.
\end{equation}
More precisely, we proved, \cf \cite[Theorem 6.7]{cg:qgravity2},
\bt
Assume $n\ge 2$ and $\so$ to be compact and let $(v,\mu)$ be a solution of the eigenvalue problem \re{1.11} with $\mu>0$, then there exist countably many solutions $(w_i,\Lam_i)$ of the implicit eigenvalue problem \re{1.15} such that
\begin{equation}
\Lam_i<\Lam_{i+1}<\cdots <0,
\end{equation}
\begin{equation}
\lim_i\Lam_i=0,
\end{equation}
and such that the functions
\begin{equation}
u_i=w_i v
\end{equation}
are solutions of the wave equation \re{1.9}. The transformed eigenfunctions
\begin{equation}
\tilde w_i(t)=w_i(\lam_i^{\frac n{4(n-1)}}t), 
\end{equation}
where
\begin{equation}
\lam_i=(-\Lam_i)^{-\frac{n-1}n},
\end{equation}
form a basis of $L^2(\R[*]_+,\Cc)$ and also of the Hilbert space $H$ defined as the completion of $C^\un_c(\R[*]_+,\Cc)$ under the norm of the scalar product
\begin{equation}
\spd w{\tilde w}_1=\int_0^\un\{\bar w'\tilde w' +t^2\bar w\tilde w\},
\end{equation}
where a prime or a dot denotes differentiation with respect to $t$. 
\et
\br
If $\so$ is not compact, then, let 
\begin{equation}
\Om\Su \so
\end{equation}
be an arbitrary relatively compact open subset of $\so$ with smooth boundary, and exactly the same results as above will be valid in the cylinder
\begin{equation}
Q(\Om)=\Om\times\R[*]_+
\end{equation}
by solving the eigenvalue problem \re{1.11} in the Sobolev space 
\begin{equation}
H^{1,2}_0(\Om)\ii H^{2,2}(\Om)
\end{equation}
and arguing further as before.  
\er

Finally, we prove under which assumptions the  lowest eigenvalue $\mu_0$ of the eigenvalue problem \re{1.11} is strictly positive. This property can also be called a mass gap. We prove the existence of a mass gap in two cases.

In the first case we prove:
\bt
Let $\so$ be compact and let $V$ satisfy
\begin{equation}
V(\F)>0\qq\tup{a.e.},
\end{equation}
then there exists $m_0$ such that for all $m\ge m_0$ the first eigenvalue $\mu_0$ of equation \re{1.11} is strictly positive with an a priori bound from below depending on the data.
\et
In the second case, we only assume $V\ge 0$ such that we may ignore the contribution of the Higgs field to the quadratic form defined by the elliptic operator in equation \re{1.11} completely, since its contribution is non-negative, and only look at the smaller operator
\begin{equation}\lae{1.27}
-(n-1)\D v-\frac n2 Rv+\al_1\frac n8 F_{ij}F^{ij} v.
\end{equation}
If we can prove that the eigenvalues of this operator are strictly positive, then the eigenvalues of equation \re{1.11} are also strictly positive.
\bt
Let $\so$ be compact, $R\le 0$,  then the smallest eigenvalue of the operator \re{1.27} is strictly positive provided either $R$ or $F_{ij}F^{ij}$ do not vanish everywhere. 
\et

\section{Definitions and notations}
The main objective of this section is to state the equations of Gau{\ss}, Codazzi,
and Weingarten for spacelike hypersurfaces $M$ in a \di {(n+1)} Lorentzian
manifold
$N$.  Geometric quantities in $N$ will be denoted by
$(\bar g_{ \al \bet}),(\riema  \al \bet \ga \de)$, etc., and those in $M$ by $(g_{ij}), 
(\riem ijkl)$, etc.. Greek indices range from $0$ to $n$ and Latin from $1$ to $n$;
the summation convention is always used. Generic coordinate systems in $N$ resp.
$M$ will be denoted by $(x^ \al)$ \resp $(\x^i)$. Covariant differentiation will
simply be indicated by indices, only in case of possible ambiguity they will be
preceded by a semicolon, i.e., for a function $u$ in $N$, $(u_ \al)$ will be the
gradient and
$(u_{ \al \bet})$ the Hessian, but e.g., the covariant derivative of the curvature
tensor will be abbreviated by $\riema  \al \bet \ga{ \de;\e}$. We also point out that
\begin{equation}
\riema  \al \bet \ga{ \de;i}=\riema  \al \bet \ga{ \de;\e}x_i^\e
\end{equation}
with obvious generalizations to other quantities.

Let $M$ be a \tit{spacelike} hypersurface, i.e., the induced metric is Riemannian,
with a differentiable normal $\n$ which is timelike.

In local coordinates, $(x^ \al)$ and $(\x^i)$, the geometric quantities of the
spacelike hypersurface $M$ are connected through the following equations
\begin{equation}\lae{01.2}
x_{ij}^ \al= h_{ij}\n^ \al
\end{equation}
the so-called \tit{Gau{\ss} formula}. Here, and also in the sequel, a covariant
derivative is always a \tit{full} tensor, i.e.

\begin{equation}
x_{ij}^ \al=x_{,ij}^ \al-\ch ijk x_k^ \al+ \cha  \bet \ga \al x_i^ \bet x_j^ \ga.
\end{equation}
The comma indicates ordinary partial derivatives.

In this implicit definition the \tit{second fundamental form} $(h_{ij})$ is taken
with respect to $\n$.

The second equation is the \tit{Weingarten equation}
\begin{equation}
\n_i^ \al=h_i^k x_k^ \al,
\end{equation}
where we remember that $\n_i^ \al$ is a full tensor.

Finally, we have the \tit{Codazzi equation}
\begin{equation}
h_{ij;k}-h_{ik;j}=\riema \al \bet \ga \de\n^ \al x_i^ \bet x_j^ \ga x_k^ \de
\end{equation}
and the \tit{Gau{\ss} equation}
\begin{equation}
\riem ijkl=- \{h_{ik}h_{jl}-h_{il}h_{jk}\} + \riema  \al \bet\ga \de x_i^ \al x_j^ \bet
x_k^ \ga x_l^ \de.
\end{equation}

Now, let us assume that $N$ is a globally hyperbolic Lorentzian manifold with a
 Cauchy surface. 
$N$ is then a topological product $I\times \mc S_0$, where $I$ is an open interval,
$\mc S_0$ is a  Riemannian manifold, and there exists a Gaussian coordinate
system
$(x^ \al)$, such that the metric in $N$ has the form 
\begin{equation}\lae{01.7}
d\bar s_N^2=e^{2\psi}\{-{dx^0}^2+\s_{ij}(x^0,x)dx^idx^j\},
\end{equation}
where $\s_{ij}$ is a Riemannian metric, $\psi$ a function on $N$, and $x$ an
abbreviation for the spacelike components $(x^i)$. 
We also assume that
the coordinate system is \tit{future oriented}, i.e., the time coordinate $x^0$
increases on future directed curves. Hence, the \tit{contravariant} timelike
vector $(\x^ \al)=(1,0,\dotsc,0)$ is future directed as is its \tit{covariant} version
$(\x_ \al)=e^{2\psi}(-1,0,\dotsc,0)$.

Let $M=\graph \fv u\so$ be a spacelike hypersurface
\begin{equation}
M=\set{(x^0,x)}{x^0=u(x),\,x\in\mc S_0},
\end{equation}
then the induced metric has the form
\begin{equation}
g_{ij}=e^{2\psi}\{-u_iu_j+\s_{ij}\}
\end{equation}
where $\s_{ij}$ is evaluated at $(u,x)$, and its inverse $(g^{ij})=(g_{ij})^{-1}$ can
be expressed as
\begin{equation}\lae{01.10}
g^{ij}=e^{-2\psi}\{\s^{ij}+\frac{u^i}{v}\frac{u^j}{v}\},
\end{equation}
where $(\s^{ij})=(\s_{ij})^{-1}$ and
\begin{equation}\lae{01.11}
\begin{aligned}
u^i&=\s^{ij}u_j\\
v^2&=1-\s^{ij}u_iu_j\equiv 1-\abs{Du}^2.
\end{aligned}
\end{equation}
Hence, $\graph u$ is spacelike if and only if $\abs{Du}<1$.

The covariant form of a normal vector of a graph looks like
\begin{equation}
(\n_ \al)=\pm v^{-1}e^{\psi}(1, -u_i).
\end{equation}
and the contravariant version is
\begin{equation}
(\n^ \al)=\mp v^{-1}e^{-\psi}(1, u^i).
\end{equation}
Thus, we have
\br Let $M$ be spacelike graph in a future oriented coordinate system. Then the
contravariant future directed normal vector has the form
\begin{equation}
(\n^ \al)=v^{-1}e^{-\psi}(1, u^i)
\end{equation}
and the past directed
\begin{equation}\lae{01.15}
(\n^ \al)=-v^{-1}e^{-\psi}(1, u^i).
\end{equation}
\er

In the Gau{\ss} formula \re{01.2} we are free to choose the future or past directed
normal, but we stipulate that we always use the past directed normal.
Look at the component $ \al=0$ in \re{01.2} and obtain in view of \re{01.15}

\begin{equation}\lae{01.16}
e^{-\psi}v^{-1}h_{ij}=-u_{ij}- \cha 000\mspace{1mu}u_iu_j- \cha 0j0
\mspace{1mu}u_i- \cha 0i0\mspace{1mu}u_j- \cha ij0.
\end{equation}
Here, the covariant derivatives are taken with respect to the induced metric of
$M$, and
\begin{equation}
-\cha ij0=e^{-\psi}\bar h_{ij},
\end{equation}
where $(\bar h_{ij})$ is the second fundamental form of the hypersurfaces
$\{x^0=\const\}$.

An easy calculation shows
\begin{equation}\lae{2.18}
\bar h_{ij}e^{-\psi}=-\tfrac{1}{2}\dot\s_{ij} -\dot\psi\s_{ij},
\end{equation}
where the dot indicates differentiation with respect to $x^0$.
 
\section{The Yang-Mills functional}

Let $N=N^{n+1}$ be a globally hyperbolic spacetime with metric $(\bar g_{\al\bet})$, $\mc G$ a compact, semi-simple, connected Lie group, $\mf g$ its Lie algebra and $E_1=(N,\mf g,\pi, \Ad(\mc G))$ the corresponding adjoint bundle with base space $N$. The Yang-Mills functional is then defined by
\begin{equation}\lae{3.1}
\begin{aligned}
J_{YM}&=\int_N-\tfrac14 F_{\mu\lam}F^{\mu\lam}=\int_N-\tfrac14 \ga_{ab}\bar g^{\mu\rho_2}\bar g^{\lam\rho_1}F^a_{\mu\rho_1}F^b_{\rho_2\lam},
\end{aligned}
\end{equation}
where $\ga_{ab}$ is the Cartan-Killing metric in $\mf g$,
\begin{equation}
F^a_{\mu\lam}=A^a_{\lam,\mu}-A^a_{\mu,\lam}+f^a_{bc}A^b_\mu A^c_\lam
\end{equation}
is the curvature of a connection
\begin{equation}
A=(A^a_\mu)
\end{equation}
in $E_1$ and
\begin{equation}\lae{3.4}
f_c=(f^a_{cb})
\end{equation}
are the structural constants of $\mf g$. The integration over $N$ is to be understood symbolically since we shall always integrate over an open precompact subset $\tilde\Om$ of $N$.

\bd\lad{3.1}
The adjoint bundle $E_1$ is vector bundle; let $E_1^*$ be the dual bundle, then we denote by 
\begin{equation}
T^{r,s}(E_1)=\underset{r}{\underbrace{E_1\otimes\cdots\otimes E_1}}\otimes\underset{s}{\underbrace{E_1^*\otimes\cdots\otimes E_1^*}}
\end{equation}
the corresponding tensor bundle and by
\begin{equation}
\C(T^{r,s}(E_1)),
\end{equation}
or more precisely,
\begin{equation}
\C(N,T^{r,s}(E_1)),
\end{equation}
the sections of the bundle, where $N$ is the base space. Especially we have
\begin{equation}
T^{1,0}(E_1)=E_1.
\end{equation}
\ed
Thus, we have
\begin{equation}
F^a_{\mu\lam}\in \C(T^{1,0}(E_1)\otimes T^{0,2}(N)).
\end{equation}

When we fix a connection $\bar A$ in $E_1$, then a general connection $A$ can be written in the form
\begin{equation}\lae{3.8} 
A^a_\mu=\bar A^a_\mu+\tilde A^a_\mu,
\end{equation}
where $\tilde A^a_\mu$ is a tensor
\begin{equation}
\tilde A^a_\mu\in \C(T^{1,0}(E_1)\otimes T^{0,1}(N)).
\end{equation}
To be absolutely precise a connection in $E_1$ is of the form
\begin{equation}
f_cA^c_\mu,
\end{equation}
where $f_c$ is defined in \re{3.4}; $A^a_\mu$ is merely a coordinate representation.
\bd
A connection $A$ of the form \re{3.8} is sometimes also denoted by $(\bar A^a_\mu,\tilde A^a_\mu)$.
\ed
Since we assume that there exists a globally defined time function $x^0$ in $N$ we may consider globally defined tensors $(\tilde A^a_\mu)$ satisfying
\begin{equation}\lae{3.11} 
\tilde A^a_0=0. 
\end{equation}
These tensors can be written in the form $(\tilde A^a_i)$ and they can be viewed as maps
\begin{equation}
(\tilde A^a_i):N\ra \mf g\otimes T^{0,1}(\so),
\end{equation}
where $\so$ is a Cauchy hypersurface of $N$, e.g.,  a coordinate slice
\begin{equation}
\so=\{x^0=\const\}.
\end{equation}
 It is well-known that the Yang-Mills Lagrangian is singular and requires a local gauge fixing when applying canonical quantization. We impose a local gauge fixing by stipulating that the connection $\bar A$ satisfies
 \begin{equation}\lae{3.14}
\bar A^a_0=0.
\end{equation}
Hence, all connections in \re{3.8} will obey this condition since we also stipulate that the tensor fields $\tilde A^a_\mu$ have vanishing temporal components as in \re{3.11}. The gauge \re{3.14} is known as the \tit{Hamilton gauge}, \cf \cite[p. 82]{faddeev:book}. However, this gauge fixing leads to the so-called Gau{\ss} constraint, since the first variation in the class of these connections will not formally yield the full Yang-Mills equations.

In a former  theorem, \cite[Theorem 2.3]{cg:uqtheory}, we  proved that the Gau{\ss} constraint does not exist and  that it suffices to consider connections of the form \re{3.8} satisfying \re{3.11} and \re{3.14} in the Yang-Mills functional $J_{YM}$:
\bt\lat{3.3}
Let  $\tilde \Om\Su N$ be open and precompact such that there exists a local trivialization of $E_1$ in $\tilde \Om$. Let $A=(\bar A^a_\mu,\tilde A^a_\mu)$ be a connection satisfying \re{3.11} and \re{3.14} in $\tilde \Om$, and suppose that the first variation of $J_{YM}$ vanishes at $A$ with respect to compact variations of $\tilde A^a_\mu$ all satisfying \re{3.11}. Then $A$ is a Yang-Mills connection, i.e., the  Yang-Mills equation
\begin{equation}
F^{a\mu}_{\lam\hp{\mu};\mu}=0
\end{equation}
is valid in $\tilde \Om$. 
\et
Let $(B_{\rho_k}(x_k))_{k\in\N}$ be a covering of $\so$ by small open balls such that each ball lies in a coordinate chart of $\so$. Then the cylinders
\begin{equation}
U_k=I\times B_{\rho_k}(x_k)
\end{equation}
are a covering of $N$ such that each $U_k$ is contractible, hence each bundle $\pi^{-1}(U_k)$ is trivial and the connection $\bar A$ can be expressed in coordinates in each $U_k$
\begin{equation}
\bar A=(\bar A^a_\mu)=f_aA^a_\mu dx^\mu.
\end{equation}
We proved in \cite[Lemma 2.5]{cg:uqtheory}:
\bl\lal{2.5}
In each cylinder $U_k$ there exists a gauge transformation $\om=\om(t,x)$ such that
\begin{equation}
\bar A^a_0(t,x)=0\qq\A\, (t,x)\in U_k
\end{equation}
after applying the gauge transformation.
\el
And in addition:
\bl\lal{2.6}
Let $U_k$, $U_l$ be overlapping cylinders and let $\om=\om(t,x)$ be a gauge transformation relating the respective representations of the connection $\bar A$ in the overlap $U_k\ii U_l$ where both representations use the Hamilton gauge, then $\om$ does not depend on $t$, i.e.,
\begin{equation}
\dot\om=0.
\end{equation}
\el

Let $E_0$ be the adjoint bundle
\begin{equation}\lae{3.23.2}
E_0=(S_0,\mf g,\pi,\Ad(\mc G))
\end{equation}
with base space $\so$, where the gauge transformations only depend on the spatial variables $x=(x^i)$. For fixed $t$ $A^a_{i,0}$ are elements of $T^{1,0}(E_0)\otimes T^{0,1}(\so)$
\begin{equation}
A^a_{i,0}\in T^{1,0}(E_0)\otimes T^{0,1}(\so),
\end{equation}
but the vector potentials $A^a_i(t,\cdot)$ are connections in $E_0$ for fixed $t$ and therefore cannot be used as independent variables, since the variables should be the components of a tensor. However, in view of the results in \rl{2.5} and \rl{2.6} the difference
\begin{equation}
\tilde A^a_i(t,\cdot)=A^a_i(t,\cdot)-\bar A^a_i(0,\cdot)\in T^{1,0}(E_0)\otimes T^{0,1}(\so).
\end{equation}
Hence, we shall define $\tilde A^a_i$ to be the independent variables such that
\begin{equation}
A^a_i=\bar A^a_i(0,\cdot)+\tilde A^a_i
\end{equation}
and we infer
\begin{equation}
A^a_{i,0}=\tilde A^a_{i,0}.
\end{equation}
In the Hamilton gauge we therefore have
\begin{equation}
F^a_{0i}=\tilde A^a_{i,0}
\end{equation}
and hence we conclude
\begin{equation}
-\tfrac14 F_{\mu\lam}F^{\mu\lam}=\tfrac12 w^{-2}g^{ij}\ga_{ab}\tilde A^a_{i,0}\tilde A^b_{j,0}-\tfrac14 F_{ij}F^{ij},
\end{equation}
where we used \re{1.1}.

Writing the density
\begin{equation}
\sqrt g=\sqrt{\det g_{ij}}
\end{equation}
in the form
\begin{equation}\lae{3.38} 
\sqrt g=\f \sqrt{\det \chi_{ij}},
\end{equation}
where $\chi$ is a fixed Riemannian metric in $S_0$, $\chi_{ij}=\chi_{ij}(x)$, such that $0<\f=\f(x,g_{ij})$ is a function, we obtain as Lagrangian function
\begin{equation}\lae{3.39}
L_{YM}=\tfrac12\ga_{ab}g^{ij}\tilde A^a_{i,0}\tilde A^b_{j,0}w^{-1}\f-\tfrac14 F_{ij}F^{ij}w\f.
\end{equation}

In order to prove a spectral resolution of the combined Hamilton operator after quantization we need to modify the Yang-Mills Lagrangian slightly. We shall call this modification process \tit{renormalization} though the renormalization is different from the usual renormalization in quantum field theory.
\br
The renormalization is necessary since the Yang-Mills energy depends quadratically on the inverse of the metric, and hence shows a wrong scaling behaviour with respect to the metric. The appropriate scaling behaviour would be linear.
\er
\bd
When we only consider metrics $\bar g_{\al\bet}$ that can be split by a given time function $x^0$, such that the Yang-Mills Lagrangian is expressed as in \re{3.39}, then we define the renormalized Lagrangian by
\begin{equation}\lae{3.33.2}
L_{YMmod}=\tfrac12\ga_{ab}g^{ij}\tilde A^a_{i,0}\tilde A^b_{j,0}w^{-1}\f^p\f-\tfrac14 F_{ij}F^{ij}w\f^p\f,
\end{equation}
where $p\in\R[]$ is real. We shall choose
\begin{equation}
p=\frac2n.
\end{equation}
An equivalent description is, that we  have replaced
\begin{equation}
F^2=F_{\al\bet}F^{\al\bet}
\end{equation}
by
\begin{equation}
F^2\f^p
\end{equation}
though this always requires that the metric is split by a time function otherwise the definition of $\f$ makes no sense.
\ed
The $\tilde A^a_i(t,\cdot)$ can be looked at to be mappings from $\so$ to $T^{1,0}(E_0)\otimes T^{0,1}(\so)$
\begin{equation}
\tilde A^a_i(t,\cdot): \so\ra T^{1,0}(E_0)\otimes T^{0,1}(\so).
\end{equation}

The fibers of $T^{1,0}(E_0)\otimes T^{0.1}(\so)$ are the tensor products
\begin{equation}
\mf g\otimes T^{0,1}_x(\so),\qq x\in \so,
\end{equation}
which are vector spaces equipped with metric
\begin{equation}
\ga_{ab}\otimes g^{ij}. 
\end{equation}
For our purposes it is more convenient to consider the fibers to be Riemannian manifolds endowed with the above metric. Let $(\zeta^p)$, $1\le p\le n_1n$, where $n_1=\dim\mf g$, be  local coordinates and
\begin{equation}
(\zeta^p)\ra \tilde A^a_i(\zeta^p)\equiv \tilde A(\zeta)
\end{equation}
be a local embedding, then the metric has the coefficients
\begin{equation}
G_{pq}=\spd{\tilde A_p}{\tilde A_q}=\ga_{ab}g^{ij}\tilde A^a_{i,p}\tilde A^b_{j,q}.
\end{equation}
Hence, the Lagrangian $L_{YMmod}$ in \re{3.33.2} can be expressed in the form
\begin{equation}
L_{YMmod}=\tfrac12G_{pq}\dot\zeta^p\dot\zeta^qw^{-1}\f^{1+\frac2n}-\tfrac14F_{ij}F^{ij}w\f^{1+\frac2n}
\end{equation}
and we deduce
\begin{equation}
\tilde\pi_p=\pde{L_{YMmod}}{\dot\zeta^p}=G_{pq}\dot\zeta^qw^{-1}\f^{1+\frac2n}
\end{equation}
yielding the Hamilton function
\begin{equation}
\begin{aligned}
&\hat H_{YMmod}=\pi_p\dot\zeta^p-L_{YMmod}\\
&=\tfrac12 G_{pq}(\dot\zeta^pw^{-1}\f^{1+\frac2n})(\dot\zeta^qw^{-1}\f^{1+\frac2n})w\f^{-({1+\frac2n})}+\tfrac14F_{ij}F^{ij}w\f^{1+\frac2n}\\
&=\tfrac 12G^{pq}\tilde\pi_p\tilde\pi_qw\f^{-({1+\frac2n})}+\tfrac14F_{ij}F^{ij}w\f^{1+\frac2n}\\
&\equiv H_{YMmod}w.
\end{aligned}
\end{equation}
Thus, the effective Hamiltonian that will enter the Hamilton constraint equation is
\begin{equation}\lae{3.51}
H_{YMmod}=\tfrac 12\f^{-({1+\frac2n})}G^{pq}\tilde\pi_p\tilde\pi_q+\tfrac14F_{ij}F^{ij}\f^{1+\frac2n}.
\end{equation}
If the Yang-Mills Lagrangian is multiplied by a coupling constant $\al_1$, then the effective Lagrangian is
\begin{equation}\lae{3.46.2}
H_{YMmod}=\al_1^{-1}\tfrac 12\f^{-({1+\frac2n})}G^{pq}\tilde\pi_p\tilde\pi_q+\al_1\tfrac14F_{ij}F^{ij}\f^{1+\frac2n}.
\end{equation}

\section{The Higgs functional} 
Let $\F$ be a scalar field, a map from $N$ to $E_1$,
\begin{equation}
\F:N\ra E_1,
\end{equation}
i.e., $\F$ is a section of $E_1$. The Higgs Lagrangian is defined by
\begin{equation}\lae{4.3}
L_H=-\tfrac12\bar g^{\al\bet}\ga_{ab}\F^a_\al\F^b_\bet-mV(\F),
\end{equation}
where $V\ge 0$ is a smooth potential and $m>0$ a constant. Given a global time function with corresponding foliation of $N$ we also consider a renormalized potential, namely, we replace $V$ by
\begin{equation}
V \f^q, \qq q=-\frac2n,
\end{equation}
such that
\begin{equation}
L_{Hmod}=-\tfrac12\bar g^{\al\bet}\ga_{ab}\F^a_\al\F^b_\bet-mV(\F)\f^q.
\end{equation}

Let us note that $V$ does not depend on the metric and hence has also the wrong scaling behaviour.

We assume for simplicity that in a local coordinate system $\F$ has real coefficients. The covariant derivatives of $\F$ are defined by a connection $A=(A^a_\mu)$ in $E_1$ 
\begin{equation}
\F^a_\mu=\F^a_{,\mu}+f^a_{cb}A^c_\mu\F^b.
\end{equation}
As in the preceding section we work in a local trivialization of $E_1$ using the Hamilton gauge, i.e.,
\begin{equation}
A^a_0=0,
\end{equation}
hence, we conclude
\begin{equation}
\F^a_0=\F^a_{,0}.
\end{equation}
Moreover, let 
\begin{equation}
\bar\F:\so\ra E_1
\end{equation}
be an arbitrary but fixed smooth section of $E_1$ depending only on $x\in \so$ and let
\begin{equation}
\tilde\F:N\ra E_1
\end{equation}
be an arbitrary smooth section, then we define
\begin{equation}
\F=\bar\F+\tilde\F
\end{equation}
to be the argument that enters in the Higgs Lagrangian but stipulate that $\tilde\F$ will the variable.

Expressing the density $g$ as in \fre{3.38} we obtain the Lagrangian
\begin{equation}
L_{Hmod}=\tfrac12 \ga_{ab}\tilde\F^a_{,0}\tilde\F^b_{,0}w^{-1}\f-\tfrac12g^{ij}\ga_{ab}\F^a_i\F^b_jw\f-mV(\F)w\f^{(1+q)}
\end{equation}
which we have to use for the Legendre transformation. Before applying the Legendre transformation we again consider the vector space $\mf g$ to be a Riemannian manifold with metric $\ga_{ab}$. The representation of $\tilde\F$ in the form $(\tilde\F^a)$ can be looked at, in a local trivialization, to be the representation of the local coordinates $(\Theta^a)$ such that the metric $\ga_{ab}$ now depends on $x$.

Let us define
\begin{equation}
p_a=\pde{L_{Hmod}}{\dot\Theta^a},\qq\dot\Theta^a=\Theta^a_{,0},
\end{equation}
then we obtain the Hamiltonian
\begin{equation}
\begin{aligned}
&\hat H_{Hmod}=p_a\dot\Theta^a-L_{Hmod}\\
&=\tfrac12\f^{-1}\ga^{ab}p_ap_b+\tfrac12 g^{ij}\ga_{ab}\F^a_i\F^b_jw\f+mV(\F)w\f^{(1+q)}\\
&\equiv H_{Hmod}w.
\end{aligned}
\end{equation}
Thus, the Hamiltonian which will enter the Hamilton constraint is
\begin{equation}\lae{4.10}
H_{Hmod}=\tfrac12\f^{-1}\ga^{ab}p_ap_b+\tfrac12 g^{ij}\ga_{ab}\F^a_i\F^b_j\f+mV(\F)\f^{(1+q)}.
\end{equation}
If the Higgs Lagrangian is multiplied by a  coupling constant $\al_2$, then 
\begin{equation}\lae{4.15.2}
H_{Hmod}=\al_2^{-1}\tfrac12\f^{-1}\ga^{ab}p_ap_b+\al_2\tfrac12 g^{ij}\ga_{ab}\F^a_i\F^b_j\f+\al_2 mV(\F)\f^{(1+q)}.
\end{equation}

\section{The Hamilton condition}
Considering the foliation given by the time function $t$ the Einstein-Hilbert functional with cosmological constant $\Lam$ can be expressed in the form
\begin{equation}
J_G=\int_a^b\int_\Om\{\tfrac14 G^{ij,kl}\dot g_{ij}\dot g_{kl}w^{-2}+(R-2\Lam)\}w\f\sqrt \chi,
\end{equation}
where we already replaced the density $\sqrt g$ by $\f \sqrt\chi$, which is due to Arnowitt, Deser and Misner \cite{adm:old}. The metric $G^{ij,kl}$ is defined by
\begin{equation}
G^{ij,kl}=\frac12(g^{ik}g^{jl}+g^{il}g^{jk})-g^{ij}g^{kl} 
\end{equation}
and its inverse is given by
\begin{equation}
G_{ij,kl}=\tfrac12\{g_{ik}g_{jk}+g_{il}g_{jk}\}-\tfrac1{n-1}g_{ij}g_{kl}.
\end{equation}
$R$ is the scalar curvature  of the metric $g_{ij}$.

The corresponding Hamiltonian $H_G$ has the form
\begin{equation}
H_G=\{\f^{-1}G_{ij,kl}\pi^{ij}\pi^{kl}-(R-2\Lam)\f\}w,
\end{equation}
\cf \cite[Section 3]{cg:qgravity2}. Hence, the Hamiltonian of the combined Lagrangian is
\begin{equation}
\mc H=H_G+H_{YMmod}+H_{Hmod},
\end{equation}
where coupling constants are already integrated in the Hamiltonians and the Hamilton equations
\begin{equation}\lae{4.6}
\dot g_{ij}=\frac{\de {\mc H}}{\de \pi^{ij}},
\end{equation}
\begin{equation}\lae{4.7}
\dot\pi^{ij}=-\frac{\de \mc H}{\de g_{ij}}
\end{equation}
are equivalent to the tangential Einstein equations 
\begin{equation}\lae{4.8}
G_{ij}+\Lam g_{ij}-T_{ij}=0,
\end{equation}
where $T_{\al\beta}$ is the stress-energy tensor comprised of the modified Yang-Mills and Higgs Lagrangians. 

The normal component of the Einstein equations 
\begin{equation}\lae{4.9}
G_{\al\bet}\nu^\al\nu^\bet-\Lam-T_{\al\bet}\nu^\al\nu^\bet=0
\end{equation}
cannot be derived from the Hamilton equations and this equation has to be stipulated as an extra condition, the so-called Hamilton condition. 

In \cite[Theorem 3.2]{cg:qgravity} we proved that any metric $(\bar g_{\al\bet})$ which splits according to \fre{1.1} satisfying \re{4.8} and \fre{4.9} also solves the full Einstein equations, i.e., it also satisfies the mixed components 
\begin{equation}
G_{0j}+\Lam g_{0j}-T_{0j}=0.
\end{equation}

The Hamilton condition is equivalent to the equation
\begin{equation}\lae{4.11}
\mc H=0
\end{equation}
and after quantization, when the quantized Hamiltonian, still denoted by $\mc H$, is a differential operator in a fiber bundle, the quantum equivalent of equation \re{4.11} is considered to be
\begin{equation}\lae{4.12}
\mc H u=0,
\end{equation}
i.e., the elements of the kernel of $\mc H$ are supposed to be the physical interesting solutions. The equation \re{4.12} is known as the Wheeler-DeWitt equation. In our former papers  \cite{cg:qgravity,cg:uqtheory} we used this approach and solved the Wheeler-DeWitt equation in a fiber bundle $E$. The Hamilton operator is then a hyperbolic operator acting only in the fibers of the bundle as a differential operator and not in the base space $\so$, which is unsatisfactory. Therefore we shall express the Hamilton condition differently. 

The foliation $M(t)$ is also the solution set of the geometric flow
\begin{equation}
\dot x=-w\nu
\end{equation}
with initial hypersurface
\begin{equation}
M_0=\so, 
\end{equation}
where $\nu$ is the past directed normal, \cf \cite[equ.\ (2.3.25)]{cg:cp}.  Let $h_{ij}$ be the second fundamental form of $M(t)$, then $\pi^{ij}$ and $h_{ij}$ are related by the equation
\begin{equation}
h_{ij}=-\f^{-1}G_{ij,kl}\pi^{kl},
\end{equation}
\cf \cite[equ. (4.6)]{cg:qgravity2}, and the second Hamilton equation
\begin{equation}
\dot\pi^{ij}=-\frac{\de \mc H}{\de g_{ij}}
\end{equation}
is equivalent to the evolution equation of the $h_{ij}$ if the tangential Einstein equations \re{4.8} are supposed to be satisfied. In \cite[Section 6]{cg:qgravity2} we used the evolution equation of the mean curvature
\begin{equation}
H=g^{ij}h_{ij}
\end{equation}
to express the Hamilton condition, i.e., we modified this equation such that it was equivalent to the Hamilton condition and we shall use this approach again in the present situation. 

We note that
\begin{equation}
\pi^{ij}=(Hg^{ij}-h^{ij})\f,
\end{equation}
and hence
\begin{equation}
(n-1)H\f=g_{ij}\pi^{ij}.
\end{equation}
We shall modify the evolution equation
\begin{equation}
\begin{aligned}
(\f^{-\frac12}g_{ij}\pi^{ij})'&= -\frac14\f^{-\frac12}g^{kl}\dot g_{kl}g_{ij}\pi^{ij}+\f^{-\frac12}\dot g_{ij}\pi^{ij}+\f^{-\frac12}g_{ij}\dot\pi^{ij}\\
&=\frac{n-1}2H^2\f^\frac12 w-2\f^{-\frac12}h_{ij}\pi^{ij}w+\f^{-\frac12}g_{ij}\dot\pi^{ij},
\end{aligned}
\end{equation}
where we used that
\begin{equation}
h_{ij}=-\frac12\dot g_{ij}w^{-1},
\end{equation}
in view of \fre{2.18}, where we emphasize that the symbol $H$ represents the mean curvature and $\mc H$ the Hamilton function. The Hamilton function is the sum of three Hamiltonians
\begin{equation}
\mc H=H_0+H_1+H_2,
\end{equation}
where $H_0$ is the gravitational, $H_1$ the renormalized Yang-Mills and $H_2$ the renormalized Higgs Hamiltonian. Thus, we infer
\begin{equation}
g_{ij}\dot\pi^{ij}=-g_{ij}\frac{\de \mc H}{\de g_{ij}}=-g_{ij}\frac{\de (H_0+H_1+H_2)}{\de g_{ij}}
\end{equation}
and we deduce further
\begin{equation}
\begin{aligned}
-g_{ij}\frac{\de  H_0}{g_{ij}}&=(\frac n2-2)\f^{-1}G_{ij,kl}\pi^{ij}\pi^{kl}w+\frac n2(R-2\Lam)\f w\\
&\q -\frac12 R\f w-(n-1)\tilde\D w\f,
\end{aligned}
\end{equation}
where the scalar curvature and the Laplacian are defined by the metric $g_{ij}$; for a proof see the proof of \cite[Theorem 3.2]{cg:qgravity2}.

Writing
\begin{equation}
H_1=\al_1^{-1}\frac12 G^{pq}\tilde\pi_p\tilde\pi_q\f^{-(1+\frac2n)}w+C_1
\end{equation}
and
\begin{equation}
H_2=\al_2^{-1}\frac12 \ga^{ab}p_ap_b \f^{-1}w+C_2
\end{equation}
we infer
\begin{equation}
-g_{ij}\frac{\de H_1}{\de g_{ij}}=\frac n2\al_1^{-1}\frac12 G^{pq}\tilde\pi_p\tilde\pi_q\f^{-(1+\frac2n)}w-g_{ij}\frac{\de C_1}{\de g_{ij}}
\end{equation}
and
\begin{equation}
-g_{ij}\frac{\de H_2}{\de g_{ij}}=\frac n2 \al_2^{-1}\frac12 \ga^{ab}p_ap_b \f^{-1}w-g_{ij}\frac{\de C_2}{\de g_{ij}}.
\end{equation}
Hence, we conclude
\begin{equation}\lae{5.29.1}
\begin{aligned}
&\msp[120](\f^{-\frac12} g_{ij}\pi^{ij})'=\\
&\q\frac1{2(n-1)}g_{ij}\pi^{ij}g_{kl}\pi^{kl}\f^\frac12w\\
&\q+\frac n2 \f^{-1}G_{ij,kl}\pi^{ij}\pi^{kl}\f^{-\frac12}w+\frac n2 (R-2\Lam)\f^\frac12 w\\
&\q -\frac12 R\f^\frac12 w-(n-1)\tilde\D w\f^\frac12\\
&\q+\frac n2\{\al_1^{-1}\frac12 G^{pq}\tilde\pi_p\tilde\pi_q\f^{-(1+\frac2n)}+\al_2^{-1}\frac12 g^{ab}p_ap_b\f^{-1}\}\f^{-\frac12}w\\
&\q -g_{ij}\{\frac{\de C_1}{\de g_{ij}}+\frac{\de C_2}{\de g_{ij}}\}\f^{-\frac12}.
\end{aligned}
\end{equation}
On the right-hand side of this evolution equation we now implement the Hamilton condition by replacing
\begin{equation}
\f^{-1}G_{ij,kl}\pi^{ij}\pi^{kl}w
\end{equation}
by
\begin{equation}
(R-2\Lam)\f w-H_1-H_2.
\end{equation}
Expressing the time derivative on the left-Hand side of  \re{5.29.1} with the help of the Poisson brackets, we finally obtain
\begin{equation}\lae{5.29}
\begin{aligned}
&\msp[120]\{\f^{-\frac12} g_{ij}\pi^{ij},\mc H\}=\\
&\q\frac1{2(n-1)}g_{ij}\pi^{ij}g_{kl}\pi^{kl}\f^\frac12w\\
&\q+\frac n2 (R-2\Lam)\f^\frac12 w-\frac n2(C_1+C_2)\f^{-\frac12}\\
&\q +\frac n2 (R-2\Lam)\f^\frac12 w-\frac12 R\f^\frac12 w-(n-1)\tilde\D w\f^\frac12\\
&\q -g_{ij}\{\frac{\de C_1}{\de g_{ij}}+\frac{\de C_2}{\de g_{ij}}\}\f^{-\frac12}.
\end{aligned}
\end{equation}
which is equivalent to the Hamilton condition if the Hamilton equations are valid.

Thus, we have proved:
\bt
Let $N=N^{n+1}$ be a globally hyperbolic spacetime and let the metric $\bar g_{\al\bet}$ be expressed as in \fre{1.1}. Then, the metric satisfies the full Einstein equations if and only if the metric is a solution of the  Hamilton equations and of the equation \re{5.29}. 
\et

\section{The quantization}
For the quantization we use a similar model as in \cite[Section 4]{cg:qgravity2}. First, we switch to the gauge $w=1$. In our previous paper we considered a bundle with base space $\so$ and fibers $F(x)$, $x\in\so$, the elements of which were the Riemannian $(g_{ij}(x))$. The fibers were equipped with the Lorentzian metric
\begin{equation}
(\f^{-1}G_{ij,kl})
\end{equation}
which, in a suitable coordinate system 
\begin{equation}
(t,\xi^A),\qq t=\f^\frac12,
\end{equation}
has the form
\begin{equation}
ds^2=-\frac{16(n-1)}ndt^2+\frac{4(n-1)}nt^2 G_{AB}d\xi^Ad\xi^B,
\end{equation}
where $G_{AB}$ is independent of $t$ and the coordinates $(t,\xi^A)$ are independent of $x$, \cf \cite[equ. (4.60)]{cg:qgravity2}.

In the present situation we consider a bundle $E$ with base space $\so$ and the fibers over $x\in\so$ are
\begin{equation}\lae{6.4}
F(x)\times (\mf g\otimes T^{01,}_x(\so))\times \mf g,
\end{equation}
where the additional components are due to the Yang-Mills fields $(\tilde A^a_i)$ and the Higgs field $(\tilde \F^a)$. Let us emphasize that the elements of the fibers are tensors and that a fixed connection $\bar A=(\bar A^a_i(x))$ and fixed Higgs field $\bar \F^a$ are used to define the connections
\begin{equation}
A^a_i=\bar A^a_i+\tilde A^a_i
\end{equation}
\resp the Higgs fields
\begin{equation}
\F^a=\bar\F^a+\tilde\F^a
\end{equation}
the terms in the Hamiltonian will depend on. After the quantization is finished and we have obtained the final equation governing the interaction of a Riemannian metric with Yang-Mills and Higgs fields, we shall choose $\tilde A^a_i=0$ and $\tilde \F^a=0$ such that only the arbitrary \tit{sections} $\bar A^a_i$ and $\bar\F^a$ are involved and not any \tit{elements} of the bundle.

The fibers in \re{6.4} are equipped with the metric
\begin{equation}\lae{6.7}
\begin{aligned}
ds^2&=-\frac{16(n-1)}ndt^2+\frac{4(n-1)}nt^2 G_{AB}d\xi^Ad\xi^B \\
&\q +t^2\al_1\tilde G_{pq}d\zeta^pd\zeta^q+t^2\al_2\ga_{ab}d\Theta^ad\Theta^b,
\end{aligned}
\end{equation}
where the metrics $\tilde G_{pq}$ and $\ga_{ab}$ are independent of $t$. The metric $G_{pq}$ in \fre{3.46.2} is related with $\tilde G_{pq}$ by
\begin{equation}
G_{pq}=t^{-\frac 4n}\tilde G_{pq}.
\end{equation}
Here, we used that a metric
\begin{equation}
g_{ij}(x)\in F(x)
\end{equation}
can be expressed in the form
\begin{equation}
g_{ij}=t^\frac4n\s_{ij},
\end{equation}
where $\s_{ij}$ is dependent of $t$ satisfying
\begin{equation}
\det\s_{ij}=\det\chi_{ij},
\end{equation}
\cf \cite[equ. (4.66)]{cg:qgravity2}.

Let us abbreviate the fiber metric in \re{6.7} by
\begin{equation}
ds^2=\bar g_{\al\bet}d\xi^\al\xi^\bet, \qq0\le \al,\bet\le n_2,
\end{equation}
such that
\begin{equation}
\xi^0=t,
\end{equation}
and let $\bar R_{\al\bet}$ be the corresponding Ricci tensor, then
\begin{equation}\lae{6.14} 
\bar R_{0\beta}=0\qq\A\,\bet
\end{equation}
as can be easily derived by introducing a conformal time
\begin{equation}
\tau=\log t
\end{equation}
such that 
\begin{equation}
\bar g_{\al\bet}=e^{2\psi} g_{\al\bet},
\end{equation}
where the coefficients $g_{\al\bet}$ are independent of $\tau$,
\begin{equation}
g_{00}=-1,
\end{equation}
and
\begin{equation}
\psi=\tau+c,\qq c=\const
\end{equation}
and using the well-known formula 
\begin{equation}
\bar R_{\alpha \beta }=R_{\alpha \beta }-(n-1)[\psi _{\alpha \beta }-\psi _\alpha
\psi _\beta ]-g_{\alpha \beta }[\Delta \psi +(n-1)\norm{D\psi}^2]
\end{equation}
connecting the Ricci tensors of conformal metrics. Norms and derivatives on the right-hand side are all with respect to the metric $g_{\al\bet}$. The index $0$ now refers to the variable $\tau$.

We can now quantize the Hamiltonian setting using the original variables $(g_{ij}, \pi^{kl},\ldots)$. We consider the bundle $E$ equipped with the metric \re{6.7} in the fibers and with the Riemannian metric $\chi$ in $\so$. Furthermore, let
\begin{equation}
C^\un_c(E)
\end{equation}
be the space of real valued smooth functions with compact support in $E$.

In the quantization process, where we choose $\hbar=1$, the variables $g_{ij}$, $\pi^{ij}$, etc.\  are then replaced by operators $\hat g_{ij}$, $\hat\pi^{ij}$, etc.\  acting in $C^\un_c(E)$ and satisfying the commutation relations
\begin{equation}
[\hat g_{ij},\hat\pi^{kl}]=i\de^{kl}_{ij},
\end{equation}
for the gravitational variables,
\begin{equation}
[\hat\zeta^p,\hat{\tilde\pi}_q]=i\de^p_q
\end{equation}
for the Yang-Mills variables, and
\begin{equation}
[\hat\theta^a,\hat p_b]=i\de^a_b
\end{equation}
for the Higgs variables, while all the other commutators vanish. These operators are realized by defining $\hat g_{ij}$ to be the multiplication operator
\begin{equation}
\hat g_{ij} u=g_{ij}u
\end{equation}
and $\hat \pi^{ij}$ to be the \tit{functional} derivative
\begin{equation}
\hat\pi^{ij}=\frac1i \dde{}{g_{ij}},
\end{equation}
i.e., if $u\in C^\un_c(E)$, then
\begin{equation}
\dde{u}{g_{ij}}
\end{equation}
is the Euler-Lagrange operator of the functional
\begin{equation}
\int_{\so}u\sqrt\chi\equiv\int_\so u.
\end{equation}
Hence, if $u$ only depends on $(x,g_{ij})$ and not on derivatives of the metric, then
\begin{equation}
\dde{u}{g_{ij}}=\pde u{g_{ij}}.
\end{equation}
The same definitions and reasonings are also valid for the other variables. Therefore, the transformed Hamiltonian $\hat {\mc H}$ can be looked at as the hyperbolic differential operator
\begin{equation}\lae{4.36}
\hat {\mc H}=-\D +C_0+C_1+C_2,
\end{equation}
where $\D$ is the Laplacian of the metric in \re{6.7} acting on functions $u\in C^\un_c(E)$ and the symbols $C_i$, $i=1,2,3$, represent the lower order terms of the respective Hamiltonians $H_0$, $H_1$ and $H_2$.

Following Dirac the Poisson brackets on the left-hand side of \fre{5.29} are replaced by $\frac1i$ times the commutators of the transformed quantities in the quantization process, since $\hbar=1$. Dropping the hats in the following to improve the readability the left-hand side of equation \re{5.29} is transformed to
\begin{equation}
i[\mc H,\f^{-\frac12}g_{ij}\pi^{ij}]=[\mc H,\f^{-\frac12}g_{ij}\frac\de{\de g_{ij}}].
\end{equation}
As we proved in \cite[equ. (6.25)]{cg:qgravity2}
\begin{equation}
\f^{-\frac12}g_{ij}\frac\de{\de g_{ij}}=\frac n4 \pde {}t
\end{equation}
when applied to functions $u$, hence
\begin{equation}
[-\D,\frac n4 \pde{}t]u=0,
\end{equation}
in view of \re{6.14}, and
\begin{equation}\lae{6.33}
[C_0+C_1+C_2,\f^{-\frac12}g_{ij}\frac\de{\de g_{ij}}]u=-(n-1)\f^{-\frac12}\tilde\D u\f -\f^{-\frac12}(\sum_{k=0}^2\frac\de{\de g_{ij}}C_k)u,
\end{equation}
\cf \cite[equ. (6.27)]{cg:qgravity2}, where $\tilde \D$ is the Laplace operator with respect to the metric $g_{ij}$. Here, we evaluate the equation \re{6.33} at an arbitrary point
\begin{equation}\lae{6.34}
(x,g_{ij},\tilde A^a_k,\tilde\F^b)\equiv (x,t,\zeta^ A)
\end{equation}
in $E$, where we used the abbreviation
\begin{equation}
(\zeta^\al)=(\zeta^0,\zeta^A)\equiv (t,\zeta^A)
\end{equation}
to denote the fiber coordinates in a local trivialization. The spatial fiber coordinates $(\zeta^A)$ are the coordinates for the fibers of the subbundle
\begin{equation}
E_1=\{t=1\}
\end{equation}
which is a Cauchy hypersurface, since the fibers of $E$ are globally hyperbolic, \cf \cite[Theorem 4.1]{cg:uqtheory}.
\br
If we consider $u$ to depend on the left-hand side of \re{6.34}, then $\tilde \D u$ has to be evaluated by applying the chain rule. However, if we consider $u$ to depend on $(x,t,\zeta^A)$, which are independent variables, then $\tilde \D u$ is the Laplacian of
\begin{equation}
u(\cdot,t,\zeta^A).
\end{equation}
We shall adopt the latter view. Indeed, after having derived the quantized version of \fre{5.29} we shall consider $u$ to depend  on $(x,t)$ and only implicitly on a fixed $\zeta^A$, i.e., on a given $(\tilde A^a_i)$ and $(\tilde \F^a)$, especially since we shall then specify
\begin{equation}
\tilde A^a_i=0\q\wed\q \tilde\F^a=0.
\end{equation}
\er

Let us now transform the right-hand side of \fre{5.29} by having in mind that $w=1$ and by multiplying all terms with $\f^\frac12$ before applying them to a function $u$. Later, when we compare the left and right-hand sides, we of course multiply the left-hand side by the same factor $\f^\frac12$.

The only non-trivial term on the right-hand side of \re{5.29} is the first one with the second derivatives. We arrange the covariant derivatives such that we obtain
\begin{equation}
-\frac1{32}\frac{n^2}{n-1} \Ddot u,
\end{equation}
where the derivatives are ordinary partial derivatives with respect to $t$, \cf the arguments in \cite[equ. (6.28)--(6.33)]{cg:qgravity2}. The other terms are trivial and we infer that the right-hand side is transformed to
\begin{equation}\lae{6.40}
\begin{aligned}
-\frac1{32}\frac{n^2}{n-1} \Ddot u-\frac n2(C_0+C_1+C_2)u-(g_{ij}\frac\de{\de g_{ij}}(C_0+C_1+C_2))u.
\end{aligned}
\end{equation}
Now, multiplying \re{6.33} by $\f^\frac12$ and observing that it equals \re{6.40}, we finally obtain the hyperbolic equation
\begin{equation}\lae{6.41}
\begin{aligned}
&\frac1{32}\frac {n^2}{n-1}\Ddot u-(n-1)\f\tilde\D u-\frac n2(R-2\Lam)\f u+\al_1\frac n8 F_{ij}F^{ij}\f^{1+\frac2n}\\
&+\al_2 \frac n4 \ga_{ab}g^{ij}\F^a_i\F^b_i\f u+\al_2\frac n2 m V(\F)\f^{1-\frac2n}u=0,
\end{aligned}
\end{equation}
where
\begin{equation}
(g_{ij},\tilde A^a_k,\tilde\F^b)
\end{equation}
are arbitrary but fixed elements of the bundle.

In \cite[equ. (6.35)--(6.37)]{cg:qgravity2} we have shown that
\begin{equation}
g_{ij}(x,t)=t^\frac4n\s_{ij}(x),
\end{equation}
where
\begin{equation}
\det \s_{ij}=\det \chi_{ij},
\end{equation}
such that
\begin{equation}
(\s_{ij},\tilde A^a_k,\tilde \F^b)
\end{equation}
belong to the subbundle $E_1$. Observing that
\begin{equation}
\tilde \D u=t^{-\frac4n}\tilde \D_{\s_{ij}}u,
\end{equation}
and
\begin{equation}
R=t^{-\frac4n}R_{\s_{ij}},
\end{equation}
where $R_{\s_{ij}}$ is the scalar curvature of the metric $\s_{ij}$, we can express \re{6.41} in the form
\begin{equation}\lae{6.48}
\begin{aligned}
&\frac1{32}\frac {n^2}{n-1}\Ddot u-(n-1)t^{2-\frac4n}\D u-\frac n2t^{2-\frac4n}Ru+\al_1\frac n8 t^{2-\frac4n}F_{ij}F^{ij}u\\
&+\al_2 \frac n4 t^{2-\frac4n}\ga_{ab}\s^{ij}\F^a_i\F^b_iu+\al_2\frac n2 m t^{2-\frac4n}V(\F)u+nt^2\Lam u=0,
\end{aligned}
\end{equation}
where we dropped the tilde from $\tilde \D u$ and where the Laplacian, the scalar curvature and the raising and lowering of indices are defined with respect to the metric $\s_{ij}$.

In \cite[Remark 6.8]{cg:qgravity2} we have proved that we may choose $\s_{ij}=\chi_{ij}$, and since $\chi_{ij}$ has been an arbitrary Riemannian metric on $\so$, we can therefore prove:
\bt
Let $(\so,\s_{ij})$ be a connected, complete, and  smooth $n$-dimensional Riemann manifold and let 
$E_0=(\so,\mf g,\pi,\Ad(\mc G))$ be the adjoint bundle defined in \fre{3.23.2}, and let
\begin{equation}
A=(A^a_i)
\end{equation}
be an arbitrary smooth connection in $E_0$, i.e., an arbitrary smooth section, and let
\begin{equation}
\F=(\F^a)
\end{equation}
be an arbitrary smooth Higgs field, then the hyperbolic equation \re{6.48} in
\begin{equation}
Q=\R[*]_+\times \so
\end{equation}
describes the quantized version of the interaction of $(\so,\s_{ij})$ with these bosonic fields.
\et
\bp 
We only have to prove that we may choose the connection $(A^a_i)$ and the Higgs field $(\F^a)$ as arbitrary smooth sections. This follows immediately by evaluating \re{6.48} at the bundle elements
\begin{equation}
\tilde A^a_i=0\q\wed\q \tilde\F^a=0,
\end{equation}
then the connection $A^a_i$ and the Higgs field $\F^a$ coincide with $\bar A^a_i$ \resp $\bar \F^a$ which are arbitrary smooth sections.
\ep
\br
If we define in $Q$ the Lorentz metric
\begin{equation}
d\bar s^2=-32\frac{n-1}{n^2}dt^2+\frac1{n-1}\s_{ij}dx^idx^j,
\end{equation}
then $Q$ is globally hyperbolic and  the operator in \re{6.48} is symmetric. If we equip $Q$ with the metric
\begin{equation}
d\bar s^2=-32\frac{n-1}{n^2}dt^2+\frac1{n-1}t^{\frac4n-2}\s_{ij}dx^idx^j,
\end{equation}
then $Q$ is also globally hyperbolic, the operator in \re{6.48} normally hyperbolic but not symmetric, and $Q$ has a big bang singularity in $t=0$ if $n\ge 3$.
\er
\bp
Since $\s_{ij}$ is complete it suffices to prove the big bang assertion. Let 
\begin{equation}
M(t)=\{x^0=t\}
\end{equation}
be the Cauchy hypersurfaces and $h_{ij}$ their second fundamental form with respect to the past directed normal, then
\begin{equation}
h_{ij}=-\frac1{2(n-1)}(t^{\frac4n-2})'\s_{ij}=p\frac1{2(n-1)}t^{-(p+1)}\s_{ij},
\end{equation}
where
\begin{equation}
p=2-\frac 4n.
\end{equation}
Hence the $M(t)$ are all umbilical. Let $H$ be the mean curvature, then
\begin{equation}
H=\frac{np}2t^{-1}.
\end{equation}
Moreover, let $\tilde R$ be the scalar curvature of the $M(t)$ and $R$ the scalar curvature of $\s_{ij}$, then
\begin{equation}
\tilde R=(n-1) t^pR
\end{equation}
and we deduce
\begin{equation}
\lim_{t\ra 0}R=0
\end{equation}
and 
\begin{equation}
\lim_{t\ra 0}H^2=\un.
\end{equation}
Hence, some sectional curvatures of the ambient metric must also get unbounded in view of the Gau{\ss} equation and the fact that the $M(t)$ are umbilical.
\ep

\section{The spectral resolution}
In case $\so$ is compact we can prove a spectral resolution for the equation \fre{6.48}, where $\Lam$ will act as an implicit eigenvalue. The proof is similar as in our previous paper \cite[Section 6]{cg:qgravity2}. First, let us consider an elliptic eigenvalue problem which can be looked at to be the stationary version of equation \re{6.48}.
\bl\lal{7.1}
Let $\so$ be compact equipped with the metric $\s_{ij}$. Then, the eigenvalue problem
\begin{equation}\lae{7.1}
\begin{aligned}
&-(n-1)\D v-\frac n2Rv+\al_1\frac n8 F_{ij}F^{ij}v\\
&+\al_2 \frac n4 \ga_{ab}\s^{ij}\F^a_i\F^b_iv+\al_2\frac n2 m V(\F)v=\mu v
\end{aligned}
\end{equation}
has countably many solutions $(v_i,\mu_i)$ such that
\begin{equation}
\mu_0<\mu_1\le \mu_2\le \cdots,
\end{equation}
\begin{equation}
\lim \mu_i=\un
\end{equation}
and
\begin{equation}
\int_\so\bar v_iv_j=\de_{ij},
\end{equation}
where now we consider complex valued functions. The solutions are smooth in $\so$ and form a basis in $L^2(\so,\Cc)$.
\el
This result is well-known. For clarification let us recall $R$ is the scalar curvature of $\s_{ij}$, and the other coefficients depend on a given smooth Yang-Mills field and a Higgs field. There is no sign condition on the potential $V$, but later, when establishing assumptions guaranteeing that
\begin{equation}
\mu_0>0,
\end{equation}
we shall require that
\begin{equation}
V\ge 0,
\end{equation}
or even
\begin{equation}\lae{7.7}
V>0\qq \tup{a.e.},
\end{equation}
i.e., $V$ is strictly positive except on a Lebesgue null set. The constant $m$ is always supposed to be non-negative.

To prove a spectral resolution of the hyperbolic equation \re{6.48} we choose an eigenfunction $v=v_i$ with positive eigenvalue $\mu=\mu_i$ and look at solutions of \re{6.48} of the form
\begin{equation}
u(x,t)=w(t) v(x).
\end{equation}
$u$ is then a solution of \re{6.48} provided $w$ satisfies the implicit eigenvalue equation
\begin{equation}\lae{7.9}
-\frac1{32}\frac{n^2}{n-1}\Ddot w-\mu t^{2-\frac4n}w-nt^2\Lam w=0,
\end{equation}
where $\Lam$ is the eigenvalue.

This eigenvalue problem we also considered in our previous paper and proved that it has countably many solutions $(w_i,\Lam_i)$ with finite energy, i.e.,
\begin{equation}
\int_0^\un\{\abs{\dot w_i}^2+(1+t^2+\mu t^{2-\frac4n})\abs{w_i}^2\}<\un.
\end{equation}
More precisely, we proved, \cf \cite[Theorem 6.7]{cg:qgravity2},
\bt\lat{7.2}
Assume $n\ge 2$ and $\so$ to be compact and let $(v,\mu)$ be a solution of the eigenvalue problem \re{7.1} with $\mu>0$, then there exist countably many solutions $(w_i,\Lam_i)$ of the implicit eigenvalue problem \re{7.9} such that
\begin{equation}
\Lam_i<\Lam_{i+1}<\cdots <0,
\end{equation}
\begin{equation}
\lim_i\Lam_i=0,
\end{equation}
and such that the functions
\begin{equation}
u_i=w_i v
\end{equation}
are solutions of the wave equations \re{6.48}. The transformed eigenfunctions
\begin{equation}
\tilde w_i(t)=w_i(\lam_i^{\frac n{4(n-1)}}t), 
\end{equation}
where
\begin{equation}
\lam_i=(-\Lam_i)^{-\frac{n-1}n},
\end{equation}
form a basis of $L^2(\R[*]_+,\Cc)$ and also of the Hilbert space $H$ defined as the completion of $C^\un_c(\R[*]_+,\Cc)$ under the norm of the scalar product
\begin{equation}
\spd w{\tilde w}_1=\int_0^\un\{\bar w'\tilde w' +t^2\bar w\tilde w\},
\end{equation}
where a prime or a dot denotes differentiation with respect to $t$. 
\et
\br
If $\so$ is not compact, then, let 
\begin{equation}
\Om\Su \so
\end{equation}
be an arbitrary relatively compact open subset of $\so$ with smooth boundary, and exactly the same results as above will be valid in the cylinder
\begin{equation}
Q(\Om)=\Om\times\R[*]_+
\end{equation}
by solving the eigenvalue problem \re{7.1} in the Sobolev space 
\begin{equation}
H^{1,2}_0(\Om)\ii H^{2,2}(\Om)
\end{equation}
and arguing further as before.  
\er

Finally, let us consider under which assumptions the lowest eigenvalue $\mu_0$ of the eigenvalue problem \re{7.1} is strictly positive. This property can also be called a mass gap. We prove the existence of a mass gap in two cases.

In the first case we assume that $V$ satisfies the condition \re{7.7}.
\bt\lat{7.4}
Let $\so$ be compact and let $V$ satisfy \re{7.7}, then there exists $m_0$ such that for all $m\ge m_0$ the first eigenvalue $\mu_0$ of equation \re{7.1} is strictly positive with an a priori bound from below depending on the data.
\et
The theorem immediately follows from a well-known compactness lemma:
\bl
Under the assumptions of the previous theorem there exists for any $\e>0$ a constant $c_\e$ such that
\begin{equation}\lae{7.20}
\int_\so \abs u^2\le \e \int_\so \abs{Du}^2+c_\e\int_\so V\abs u^2\qq\A\, u\in C^1(\so).
\end{equation}
\el
\bp 
We prove the estimate \re{7.20} in he Sobolev space $H^{1,2}(\so)$ instead of $C^1(\so)$, since this is the appropriate function space, and argue by contradiction.

If the estimate \re{7.20} would be false, then there would exist $\e>0$ and a sequence of functions
\begin{equation}
u_k\in H^{1,2}(\so)
\end{equation}
such that
\begin{equation}
\int_\so \abs{u_k}^2>\e \int_\so \abs{Du_k}^2+k\int_\so V\abs{u_k}^2.
\end{equation}
Without loss of generality we may assume
\begin{equation}
\int_\so\abs{u_k}^2=1.
\end{equation}
Hence, the $u_k$ are bounded in $H^{1,2}(\so)$ and a subsequence, not relabeled, will weakly converge in $H^{1,2}(\so)$ to a function $u$ such that
\begin{equation}
u_k\ra u\qq\tup{in}\q L^2(\so),
\end{equation}
since the embedding from $H^{1,2}(\so)$ into $L^2(\so)$ is compact, and we would deduce
\begin{equation}
\int_\so \abs u^2=1
\end{equation}
and also
\begin{equation}
\int_\so V\abs u^2=0,
\end{equation}
a contradiction.
\ep

In the second case, we only assume $V\ge 0$ such that we may ignore the contribution of the Higgs field to the quadratic form defined by the elliptic operator in equation \re{7.1} completely, since its contribution is non-negative, and only look at the smaller operator
\begin{equation}\lae{7.27}
-(n-1)\D v-\frac n2 Rv+\al_1\frac n8 F_{ij}F^{ij} v.
\end{equation}
If we can prove that the eigenvalues of this operator are strictly positive, then the eigenvalues of equation \re{7.1} are also strictly positive.
\bt
Let $\so$ be compact, $R\le 0$,  then the smallest eigenvalue of the operator \re{7.27} is strictly positive provided either $R$ or $F_{ij}F^{ij}$ do not vanish everywhere.
\et
\bp
Under the assumptions the eigenvalues are always non-negative and the spectral resolution described in \rl{7.1} is valid. Therefore, assume that $\mu_0=0$ and let $u$ be a corresponding eigenfunction, then
\begin{equation}
0=\int_\so\abs{Du}^2-\frac n2\int_\so R\abs u^2+\al_1\frac n8\int_\so F_{ij}F^{ij} \abs u^2.
\end{equation}
Hence, each of the integrals will vanish and we conclude that
\begin{equation}
u=\const
\end{equation}
and
\begin{equation}
-R+ F_{ij}F^{ij}=0,
\end{equation}
contradicting the assumptions.
\ep
\br
The eigenfunctions $v$ of the eigenvalue problem \re{7.1} certainly have a mass if the assumptions of \rt{7.4} are satisfied and the eigenfunctions $w$ of \re{7.9} have positive energy independently of any Yang-Mills field, only because of the interaction of gravity with the scalar field. We therefore believe that the eigenfunctions $v$ could be responsible for the dark matter and the corresponding eigenfunctions $w$ for the dark energy, where
\begin{equation}
u=wv
\end{equation}
has to be a solution of the hyperbolic equation \fre{6.48}.
\er

%\backmatter
%\includepdf[pages=-]{/Users/claus/Documents/Scanned-Documents/}
\bibliographystyle{hamsplain}
%\bibliography{mrabbrev,publications}
\providecommand{\bysame}{\leavevmode\hbox to3em{\hrulefill}\thinspace}
\providecommand{\href}[2]{#2}

%\listoffigures

%\cleardoublepage

%\thispagestyle{empty}
%\closegraphsfile
\end{document}